\documentstyle[11pt]{article}
\begin{document}
\begin{flushright}SJSU/TP-98-17\\ September 1998\end{flushright}
\vspace{1.7in}
\begin{center}\Large{\bf Comment on \\ ``Consistency, amplitudes,
        and probabilities \\ in quantum theory'' \\} 
\vspace{1cm}
\normalsize\ J. Finkelstein\footnote[1]{
        Participating Guest, Lawrence Berkeley National Laboratory\\
        \hspace*{\parindent}\hspace*{1em}
        e-mail: JLFinkelstein@lbl.gov}\\
        Department of Physics\\
        San Jos\'{e} State University\\San Jos\'{e}, CA 95192, U.S.A
\end{center}
\begin{abstract}
In a recent article [Phys.\ Rev.\ A {\bf 57}, 1572 (1998)] Caticha has
concluded that ``nonlinear variants of quantum mechanics are inconsistent.''
In this note we identify what it is that nonlinear quantum theories
have been shown to be inconsistent with. 
\end{abstract}
\newpage
In a recent article (ref. \cite{C1}, hereinafter denoted by [C]), Caticha has
presented a program for justifying the standard formalism of quantum
theory, based on very general assumptions about amplitudes.  
(See also ref. \cite{C2}) One of the conclusions of this program
is that the time evolution of the quantum-mechanical wave-function 
must necessarily be linear.
To quote from the abstract of [C], an argument of which ``the crucial
ingredient is a very natural consistency constraint: if there are two
different ways to compute the amplitude for a given process, the two 
answers must agree'' leads to the conclusion that ``the Schr\"{o}dinger
equation must be linear: i.e., nonlinear variants of quantum mechanics
are inconsistent.''

There has been considerable work done on nonlinear quantum theories
\cite{nlqt}, and while these theories have been criticized (for example)
as being incompatible with relativistic causality \cite{ftl}, it would be 
surprising if they could be completely ruled out by a simple requirement 
of self-consistency.
The purpose of this note is to identify what it is that [C] has shown
nonlinear theories to be inconsistent with.

Let us consider the example from [C] of a double-slit experiment: a beam
of particles is incident on two slits, denoted as $a$ and $a'$, either of
which may be open or closed.  The transmitted beam is detected at some
fixed point on a screen.  The program of [C] begins by imagining that
a complex number $\phi$ is assigned to each of the possible setups:
$\phi(a)$ is the number assigned to the setup in which only slit $a$
is open, $\phi(a')$ to the setup in which only $a'$ is open, and
$\phi(a\vee a')$ to the setup in which both slits are open.  [C] then
goes on to establish, as indicated briefly below, that for such an
assignment, there exists a function $\xi (\phi)$ which satisfies the
following additivity property:
\begin{equation}
  \xi (\phi (a \vee a')) = \xi (\phi (a)) + \xi (\phi (a')).
\end{equation}
This equation also appears as eq.\ (C17) (where eq.\ (Cn) 
denotes eq.\ (n) of [C]).  One can then replace the numbers $\phi$ by
the numbers $\xi (\phi)$, and write $\xi(a\vee a')=\xi(a)+\xi(a')$.
The complex numbers $\xi$ are then identified 
(up to a multiplicative constant)
as the amplitudes for the respective processes.  The existence \cite{triv} 
of a function $\xi$ satisfying this additivity condition 
(eq. (1) or (C17)) is an important step in establishing that the 
wave function obeys the linear Schr\"{o}dinger equation, as in eq. (C30).

Although the program of [C] aims to derive the meaning, and eventually
the linearity, of the wave function, it is nevertheless instructive to
imagine that we had a wave function which propagates non-linearly, and
to see how this runs afoul of [C]'s program.  So let us say that, at the
time $t$ when our particle arrives at the slits, its wave function is,
in a hopefully-obvious notation,
\begin{equation}
|\Psi(t)\rangle = \alpha | a \rangle +\alpha ' | a' \rangle.
\end {equation}
and that the effect of closing a slit is to project out that part of the
wave function (so that if for example slit $a'$ is closed, the wave
function is as in eq.\ (2) but with the number $\alpha '$ replaced by $0$).
The particle is to be detected at time $t_f$ at some point on the screen
called $x_f$, so let us define 
$\gamma := \langle x_{f} | \Psi(t_{f}) \rangle $.
Of course if the time evolution of $|\Psi \rangle$ were linear, 
$\gamma$ would be a linear function of
$\alpha$ and $\alpha '$, but let us suppose instead that
$\gamma = (\alpha + \alpha ')^2$.  This suggests that we attempt to begin
the program of [C] with the following assignment of complex numbers:
\begin{eqnarray}
   \phi (a \vee a') & = & (\alpha + \alpha ')^2 \nonumber \\
   \phi (a )  & = & (\alpha)^2 \nonumber \\
   \phi (a' )  & = & (\alpha ')^2
\end{eqnarray}
and see what goes wrong.

What happens is that there is no non-trivial function $\xi$ satisfying
eq.\ (1).  To see this, note that eqs.\ (1) and (3) imply
\begin{equation}
  \xi ((\alpha + \alpha ')^2) = \xi(\alpha ^2) + \xi(\alpha '^2).
\end{equation}
Now if in eq.\ (4) we choose $\alpha ' = \alpha $, we get
\begin{equation}
   \xi (4\alpha^2) = 2 \xi (\alpha^2),
\end{equation}
while if in eq.\ (4) we choose $\alpha ' = -\alpha $, we get
\begin{equation} 
  \xi (0) = 2 \xi (\alpha^2).
\end{equation}
Eqs.\ (5) and (6) imply that $\xi =$ constant, and then eq.\ (4)
requires $\xi = 0$.

Since with the choice made in eq.\ (3) there is no (non-trivial)
solution to eq.\ (1), we should examine under what conditions 
eq.\ (1) (i.e., eq.\ (C17)) was derived.  In [C] it is required that
the numbers $\phi$ form what is called a ``representation''.  
This requirement is expressed by eqs\ (C13) and (C19); for simplicity let
us restrict our attention to (C13), which asserts that there is a 
function $S(x,y)$ such that
\begin{equation}
    \phi (a \vee a') = S(\phi(a),\phi(a')).
\end{equation}
The consistency constraint is then stated in terms of $S$, applied to the
generalization to the case of three slits:  calling the slits $a$, $a'$,
and $a''$, the combination of all three slits could be considered either
as the combination of $a$ with $a'$, which then is combined with $a''$,
or alternatively, as $a$ combined with the combination of $a'$ and $a''$
(i.e., the associativity relation $[(a\vee a')\vee a''] =
[a\vee (a'\vee a'')]$).  Then the requirement that the value of
$\phi(a\vee a'\vee a'')$ be the same if calculated in different ways
leads, by repeated application of eq.\ (7), to
\begin{equation}
   S(S(\phi(a),\phi(a')),\phi(a''))=S(\phi(a),S(\phi(a'),\phi(a''))).
\end{equation}
This equation (which is equivalent to eq.\ (C15)) is the consistency
constraint from which eq.\ (1) (eq.\ (C17)), and ultimately the 
Schr\"{o}dinger equation (C30) is derived. 

If we extend the assignment written in eq.\ (3) in the obvious way
to the case of three slits, we see that the first place this extended
assignment fails to follow the program of [C] is {\em not} the 
consistency condition eq.\ (8), but rather in the demand that the
numbers form a ``representation'', eq.\ (7).  This equation demands that
if you know the values of $\phi(a)$ and of $\phi(a')$, you have enough
information to calculate the value of $\phi(a\vee a')$.  But with the
assignment (3), that is not true: if all you know are the values of
$\alpha^2$ and of $\alpha '^2$, you do not have enough information to
calculate the value of $(\alpha + \alpha ')^2$, since you do not know
the relative sign of $\alpha $ and $\alpha '$.

This situation is analogous to the calculation of probabilities
in standard quantum theory; if all you know are the detection 
probabilities when
slits $a$ and $a'$ are open individually, you do not have enough
information to calculate the probability when both slits are open,
because the probabilities are non-linear functions of the amplitudes.
Likewise, in a non-linear theory such as we are imagining here,
knowing the amplitudes when the slits are open individually does
not allow you to calculate the amplitude when both slits are open.
Thus the demand that the amplitudes form a representation fails for
this sort of theory.

[C] has shown that, as long as the numbers $\phi$ form a 
``representation'' (as in eq.\ (C13)), they might as well be taken to be 
additive, because
they could always be replaced by the numbers $\xi$ which {\em are}
additive.  This fact (together with the other part of the representation
requirement, eq. (C19)) then leads to the demonstrated linearity of
the Schr\"{o}dinger equation.  Thus it is the demand that the amplitudes
form a ``representation''  with which nonlinear theories are inconsistent.

\vspace{1cm}
{\bf Acknowledgement:}
I have benefitted greatly from correspondence with Ariel Caticha.
I  acknowledge also the hospitality of the
Lawrence Berkeley National Laboratory.


\begin{thebibliography}{9}
\bibitem{triv} Eq.\ (1) certainly has the trivial solution 
              [$\xi(\phi)=0$ for all $\phi$], but the program of [C] 
              needs a non-trivial solution of eq.\ (1) (for example, when 
              the amplitude is assumed to be normalized in eq.\ (C33)).
\end{thebibliography}
\end{document}